%% file: paper.tex
\documentclass[a4paper, conference]{IEEEtran}

\usepackage[T1]{fontenc}
\usepackage[utf8]{inputenc}

\usepackage[cmintegrals]{newtxmath}
\usepackage{bm}
\usepackage{cite}
\usepackage{color}
\usepackage{graphicx}
\usepackage{glossaries}
\usepackage{booktabs}
\usepackage{dcolumn}
\usepackage{xurl}
\usepackage{siunitx}
\usepackage{tablefootnote}
\DeclareSIUnit \belc {Bc}
\DeclareSIUnit{\belmilliwatt}{Bm}
\DeclareSIUnit{\dBm}{\deci\belmilliwatt}
\DeclareSIUnit{\dBc}{\deci\belc}
\DeclareSIUnit{\sample}{Sa}
\DeclareSIUnit{\ppb}{ppb}
\DeclareSIUnit{\g}{g}

\usepackage{hyperref}
\hypersetup{
	pdfauthor={Maxmilian Engehardt, Carsten Andrich, Daniel Stanko, Alexander Ihlow, Markus Landmann},
	pdftitle={A Road-Mobile GNSS-Disciplined Oscillator for Accurate Synchronization of Vehicular Microwave Measurements},
}

\usepackage{orcidlink}

\usepackage{placeins}

\usepackage{tikz}
\usetikzlibrary{positioning, matrix}
\usetikzlibrary{shapes}
\usetikzlibrary{decorations.markings, spy, fit, patterns, arrows}
\usetikzlibrary{intersections}
\usepackage{circuitikz}
\tikzset{>=latex}
\pgfdeclarelayer{bg}
\pgfdeclarelayer{fg}
\pgfsetlayers{bg,main,fg}

\usepackage{amsopn}
\usepackage{mathtools}

\usepackage{pgfplots}
\usepgfplotslibrary{fillbetween}
\pgfplotsset{compat=1.16}

\makeglossaries

\newacronym{SDR}{SDR}{software-defined radio}
\newacronym{DAC}{DAC}{digital-to-analog converter}
\newacronym{OCXO}{OCXO}{oven-controlled crystal oscillator}
\newacronym{EFC}{EFC}{electronic frequency control}
\newacronym{ENOB}{ENOB}{effective number of bits}
\newacronym{MCU}{MCU}{microcontroller unit}
\newacronym{PSRR}{PSRR}{power supply rejection ratio}
\newacronym{SPI}{SPI}{serial peripheral interface}
\newacronym{TDC}{TDC}{time-to-digital converter}
\newacronym{PCB}{PCB}{printed circuit board}
\newacronym{GNSS}{GNSS}{global navigation satellite system}
\newacronym{GNSSDO}{GNSSDO}{GNSS-disciplined oscillator}
\newacronym{FPU}{FPU}{floating point unit}
\newacronym{ESD}{ESD}{electrostatic discharge}
\newacronym{IMU}{IMU}{inertial measurement unit}
\newacronym{RTCM}{RTCM}{radio technical commission for maritime services}
\newacronym{UTC}{UTC}{coordinated universal time}
\newacronym{SYSREF}{SYSREF}{system synchronization reference signal}
\newacronym{1PPS}{1PPS}{1 pulse per second}
\newacronym{LTE}{LTE}{long term evolution}
\newacronym{LVCMOS}{LVCMOS}{low voltage complementary metal oxide semiconductor}
\newacronym{MEMS}{MEMS}{micro-electromechanical system}
\newacronym{VCO}{VCO}{voltage-controlled oscillator}

\pgfplotsset{
height=6cm,
width=8.5cm,
}

\newcommand*\circled[1]{\tikz[baseline=(char.base)]{
            \node[shape=circle,draw,inner sep=2pt] (char) {#1};}}

\begin{document}

\title{A Road-Mobile GNSS-Disciplined Oscillator\\for Accurate Synchronization\\of Vehicular Microwave Measurements}

\author{
Maximilian~Engelhardt\IEEEauthorrefmark{1}\raisebox{.5ex}{\orcidlink{0009-0002-9440-8615}}, Carsten~Andrich\IEEEauthorrefmark{2}\raisebox{.5ex}{\orcidlink{0000-0002-4795-3517}}, 
Daniel~Stanko\IEEEauthorrefmark{1}\IEEEauthorrefmark{2}\raisebox{.5ex}{\orcidlink{0009-0007-7292-8552}}, 
Alexander~Ihlow\IEEEauthorrefmark{2}\raisebox{.5ex}{\orcidlink{0000-0002-9714-4881}},  
Markus~Landmann\IEEEauthorrefmark{1}\IEEEauthorrefmark{2}\raisebox{.5ex}{\orcidlink{0009-0001-4270-0342}} \\
\\
\IEEEauthorblockA{\IEEEauthorrefmark{1}Fraunhofer Institute for Integrated Circuits IIS, Ilmenau, Germany}%
\IEEEauthorblockA{\IEEEauthorrefmark{2}Technische Universität Ilmenau, Institute of Information Technology, Ilmenau, Germany}
}

\maketitle

\begin{abstract}

Precise synchronization is essential in various technical disciplines, being especially challenging in mobile scenarios.
Unfortunately, state-of-the-art global navigation satellite system (GNSS) disciplined oscillators (GNSSDOs) are designed and optimized for stationary operation.
We present a novel solution that is optimized for mobile use from the ground up.
The centerpiece is a precise oven-controlled crystal oscillator (OCXO) that is optimized for low sensitivity to dynamic accelerations.
A state-of-the-art GNSS timing module is used to discipline it.
We evaluate the system by comparing it with state-of-the-art test equipment in a real-world test drive through diverse environments.
After compensating for the stationary offset, the state-of-the-art devices deviated by up to $\mathbf{\SI{2315}{\nano\second}}$, while with our devices, the deviation never exceeded $\mathbf{\SI{22.6}{\nano\second}}$.
It is evident that the devices designed for laboratory use perform inadequately in mobile operation and that our novel solution enables a significant leap in accuracy.

\end{abstract}


\section{Introduction}
Precise synchronization is a fundamental requirement for a wide range of RF measurement applications.
With radio waves propagating at the speed of light, a timing offset of \SI{10}{\nano\second} causes a spatial measurement error of \SI{3}{\meter}.
The conventional solution for synchronizing multiple nodes in distributed measurement setups is to use \glspl{GNSSDO}.
These combine a precise oscillator, for example, an \gls{OCXO}, with a satellite receiver to boost its long-term stability.
Unfortunately, state-of-the-art devices are designed and optimized for stationary operation, i.e., fixed position and no dynamic accelerations.
In mobile applications, the performance of these devices is severely degraded, which has presented us with difficulties in past measurement campaigns.
To solve this challenge, we present a novel \gls{GNSSDO} that is designed from the ground up for use in mobile applications, using components optimized for this environment.

\section{State of the Art}
As an example of a widespread time and frequency standard for laboratory use, we present the \textit{FS740} by \textit{Stanford Research Systems}~\cite{SOTA-FS740}, which is also used as a benchmark in \autoref{sec:measurement-results}.
Essentially, this is a \gls{GNSSDO} that is based on an \gls{OCXO}, with an optional rubidium reference.
It features a highly flexible front-end that allows both to generate a wide range of signals and to time-tag external events.
However, the device is not designed for mobile use and the \gls{GNSS} module employed has limitations:
It does not support multi-band operation and the exchange of correction data for differential operation.

The paper~\cite{SOTA-MEMS} presents a \gls{GNSSDO} optimized for mobile use.
It employs a \gls{MEMS} oscillator, making it a very cost-effective device.
However, it uses a \gls{GNSS} receiver that is optimized for positioning, not timing applications (\textit{ZED-F9P}).
The paper provides an in-depth analysis of the measured signals, but does not validate the performance in real-world mobile use.

Finally, we would like to highlight the synchronization solution for microwave measurements presented in~\cite{SOTA-KIT}.
The focus there is not on distributed synchronization of multiple nodes, but on how to ensure phase-coherent operation of multiple measuring devices in a single node.

\section{Concept}
Each node in our distributed measurement setup is equipped with a \gls{GNSSDO}.
This provides the necessary synchronization signals to all measurement devices in the node.
Here, we use the \gls{LVCMOS} standard (\SI{3.3}{\volt} logic level) for all signals.

\subsection{Reference Clock Signal}
A conventional way to synchronize measurement setups is to distribute a 10 MHz sine wave as reference signal.
However, this has several disadvantages, which arise in the downstream measurement equipment:
These devices employ PLLs to generate the necessary frequencies in the gigahertz range.
If such a circuit multiplies the input frequency by a factor of $N$, the phase noise is also amplified by a factor of $N$~\cite{buch}.\footnote{This only applies within the bandwidth of the PLL loop filter -- further away from the carrier, the phase noise is dominated by the \gls{VCO} and therefore independent of the quality of the reference signal~\cite{buch}.}
It follows that reducing this factor $N$ improves the synthesized clock.
This is achieved by selecting a higher reference clock frequency.
Another problem arises from the relatively low slew rate of a sine wave, which increases the uncertainty in detecting its zero crossings.
Therefore, our \gls{GNSSDO} is designed to produce a \SI{100}{\mega\hertz} square wave as reference clock.
However, it features optional integer division to generate lower output frequencies, especially the more conventional \SI{10}{\mega\hertz}.


\subsection{Pulse Output}\label{subsec:pulse-output}
A reference clock alone is not sufficient to synchronize the connected devices, i.e., to align their internal time bases.
This requires an additional pulse signal that resolves the periodicity of the reference clock.
In our application, this pulse signal is retimed by the connected measuring devices using the reference clock.
In other words, the pulse is not used as a time reference itself, but only marks the following edge of the reference clock.
The pulse outputs can be configured to not only provide the typical \gls{1PPS} signal but also to generate a slow clock, which can be utilized as \gls{SYSREF} for synchronizing subsequent devices in accordance with JESD204~\cite{JESD204D}.

\subsection{Pulse Input}
In addition to the outputs, the \gls{GNSSDO} also has one pulse input that can be used to time-tag external events.
This is useful not only for the actual application but also for the validation described in Section~\ref{sec:measurement-results}.

\section{Implementation}

\begin{figure}[h!]
\centering
\scalebox{0.8}{\input{figures/BlockDiagram}}
\caption{
Simplified block diagram of our \gls{GNSSDO} setup.
The clock source is the specifically acceleration-insensitive \gls{OCXO}, whose drift relative to GNSS time is measured by a time-tagging unit.
From this, a digital control algorithm in the \gls{MCU} generates an actuating signal to discipline the oscillator.
Five reference clock and pulse outputs are provided for synchronizing the connected measurement hardware.
}
\label{fig:BlockDiagram}
\end{figure}
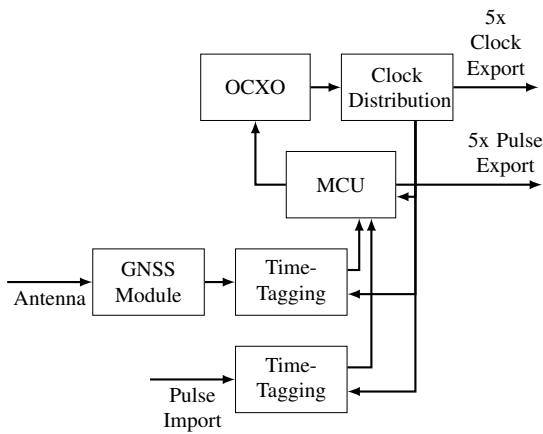

The block diagram in Figure~\ref{fig:BlockDiagram} provides an overview of our design, with the individual components described in the following.

\subsection{Oven Controlled Crystal Oscillator}

At the heart of the device is a precision \gls{OCXO}, whereby the \textit{O-40-ULGS-100M} manufactured by \textit{KVG GmbH} was selected here.
The key advantages of this model are its low phase noise  (\SI{-165}{\dBc} at \SI{1}{\kilo\hertz} offset) and its very low acceleration sensitivity of only \SI{0.2}{ppb/g}.
This makes it an ideal choice for precision mobile applications.
It also supports \gls{EFC}, which allows the frequency to be adjusted within a range of $\pm \SI{2}{ppm}$.
This is necessary to implement the control loop that forms the basis of a \gls{GNSSDO}.

The \gls{EFC} of the \gls{OCXO} is actuated via a voltage signal in the range of \qtyrange{0}{10}{\volt}.
Since the control loop is implemented digitally in the \gls{MCU}, a \gls{DAC} stage is necessary.
As the \gls{OCXO} effectively is a frequency modulator, amplitude noise at the \gls{EFC} input is converted into phase noise in the output clock.\footnote{Since the modulation depth is very low here ($\ll \SI{1}{ppm}$), this relationship can be described mathematically using the narrowband modulation theorem.}
Therefore, a low-noise design of the \gls{DAC} stage is necessary to optimize \gls{OCXO} performance.


The core of the \gls{DAC} stage is an \textit{AD5542}, which offers a resolution of 16 bits and is connected to the \gls{MCU} via \gls{SPI}.
This is followed by an active low-pass stage, which also provides the necessary gain to cover the voltage range of the \gls{EFC}.
With its low cut-off frequency of \SI{100}{\hertz}, it reduces the bandwidth of the thermal noise.
%
%
As an additional benefit, this allows the use of dithering, i.e., rapidly switching between two adjacent quantization levels of the DAC, to reduce quantization noise.
A simulation determined that the \gls{EFC} signal has an RMS noise amplitude of \SI{2.75}{\micro\volt}, corresponding to an \gls{ENOB} of 21.8 bits.

\subsection{GNSS Receiver}
The \gls{GNSS} receiver selected is the \textit{u-blox ZED-F9T}, a module optimized for precision timing applications.
It supports multi-band operation, reducing the impact of multi-path propagation and ionospheric effects.
It also features differential timing mode, whereby the precision is further enhanced by exchanging correction messages with neighboring stations.
This is realized by setting up one module as a stationary base station with an unrestricted view of the sky.
It produces correction messages in the \gls{RTCM} 3.3 format, which are then distributed to the mobile modules via internet protocol messages.
This mode of operation reduces the time deviations between the individual nodes – exactly what is needed in a distributed measurement campaign.

For easy integration, the manufacturer offers the \textit{RCB-F9T}, a board that combines the receiver chip with some support circuitry.
It connects to the \gls{PCB} of the \gls{GNSSDO} developed here via a pin header.
Interaction takes place via a UART interface, on which digital messages are exchanged with the \gls{MCU}, and a pulse output, which is connected to a time-tagging unit as described in Sec.~\ref{subsec:time-tagging}.
Instead of the conventional one pulse per second, the module is configured here to produce one pulse every \SI{100}{\milli\second}.
This allows averaging the time-tagging result over multiple pulses, thus reducing jitter.

\subsection{MCU and High-Resolution Timer}
One of the device's key tasks is to output precisely timed pulses.
To minimize hardware complexity, an \gls{MCU} that provides this feature natively was selected.
The \textit{STM32G474} contains a so-called high-resolution timer that fulfills this requirement:
It allows each clock cycle to be divided into 32 sub-steps, so that pulses can be generated with a resolution of \SI{312.5}{\pico\second}.
With this feature, the phase relationship between the reference clock and the output pulses can be adjusted in small steps.
This timer block is divided into six independent subunits, so that the five pulse outputs of the device can be individually controlled.
This provides additional flexibility; for example, outputs can be switched between \gls{1PPS} and JESD204 \gls{SYSREF} modes separately.

Other useful features of the \gls{MCU} are a hardware \gls{FPU} and a wide range of communication interfaces, including high-speed USB, which is used here to connect the device to the host.


\subsection{Time-Tagging}\label{subsec:time-tagging}
The device requires two time-tagging units: one for pulses from the \gls{GNSS} module and another for the external pulse input.
The aforementioned \gls{MCU} timer has several capture units that store the timer value at which external events occur.
However, its high resolution feature is not available here; only whole clock periods can be counted, giving a resolution of \SI{10}{\nano\second}.
To address this limitation, an additional \gls{TDC} is used here.
The \textit{TDC7200} from TI allows the time intervals occurring here to be measured with a standard deviation of only \SI{35}{\pico\second}.
Since its maximum input frequency is \SI{16}{\mega\hertz}, it cannot be directly connected to the \gls{OCXO} clock.
Instead, like the \gls{MCU}, it is supplied with a \SI{10}{\mega\hertz} clock.


\subsection{Clock Distribution and Outputs}\label{subsec:clock-distribution}

The clock distribution module is responsible for generating two groups of signals from the \gls{OCXO} clock:
\begin{itemize}
    \item \SI{10}{\mega\hertz} clocks to drive the \gls{MCU} and the \glspl{TDC}, and 
    \item The reference clock signals to be exported, supporting multiple frequencies: 100, 50, 20, and \SI{10}{\mega\hertz}
\end{itemize}
For this task, the \textit{LMK01801} from \textit{Texas Instruments} was chosen as it offers the required number of outputs and is optimized for low noise and jitter.
It contains the necessary flexible clock dividers, which can be configured via \gls{SPI}.


The clock outputs of the distribution chip, as well as the pulse outputs of the \gls{MCU}, are single-ended \gls{LVCMOS} signals.
An output buffer stage provides the current drive capability to interface with \SI{50}{\ohm} loads and to ensure short-circuit protection.

\subsection{Further Hardware Aspects}
Since the performance of the \gls{OCXO} depends on its supply voltage, this is especially critical.
For this reason, the high-precision \textit{LT3045} linear regulator was selected.
In addition to a particularly high \gls{PSRR} of $ > \SI{76}{\deci\bel}$, this also offers very low output noise.

Furthermore, an \gls{IMU} was installed to record accelerations and rotations acting on the device.
This is initially only used as a monitoring feature, but could also become the basis for acceleration compensation algorithms in the future.


The devices are mounted in standard 19-inch housings to integrate them into a rack with our measurement equipment.
However, the hardware itself would also allow for a significantly smaller form factor, as the \gls{PCB} measures only \qtyproduct[product-units = single]{20 x 11}{\centi\metre}.


\section{Measurement Results}\label{sec:measurement-results}

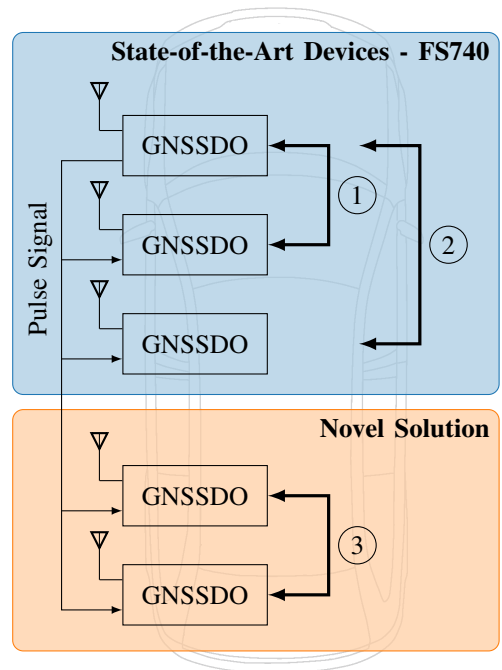
\begin{figure}[h]
\centering
\scalebox{1}{\input{figures/MeasurementSetup}}
\caption{
Measurement setup: 
Several \glspl{GNSSDO} are mounted in a single vehicle, one of which regularly emits pulses that are captured by the time-tagging inputs of the others. 
The relative timing error $\Delta t$ between two devices is determined by subtracting the recorded timestamps.}
\label{fig:MeasurementSetup}
\end{figure}

The key performance metric for our application is the relative timing error between the individual moving nodes.
In contrast, the absolute error with respect to a ground truth, for example \gls{UTC}, is less relevant.
In line with this, the deviation between multiple devices is measured here.
The simplest and most accurate way to achieve this is for the participating devices to exchange time pulses via cables.
However, this requires all equipment to be installed in the same vehicle.
To ensure that reality is represented as accurately as possible, the devices were installed in different orientations.
This ensures that the occurring accelerations act differently on the devices.
The \gls{GNSS} antennas were also distributed as widely as possible across the vehicle body.

The measurement setup is shown in Figure~\ref{fig:MeasurementSetup}.
It incorporates two novel \glspl{GNSSDO} and three FS740 devices, enabling comparison with the state-of-the-art.
These devices are equipped with the optional rubidium time base and have been configured for mobile operation.
One FS740 emits one pulse every \SI{100}{\milli\second}, which is distributed via length-matched coaxial cables to the time-tagging inputs of all other devices.
Each device records when it observed this pulse relative to its own time base.
The relative timing error of two devices can then be determined by calculating the differences between these timestamps.
These measurements are then corrected for constant offsets of the devices, which had been determined beforehand in a laboratory test.
All devices were allowed to warm up for \SI{14}{\hour} in a garage and further \SI{2}{\hour} under the open sky to settle.

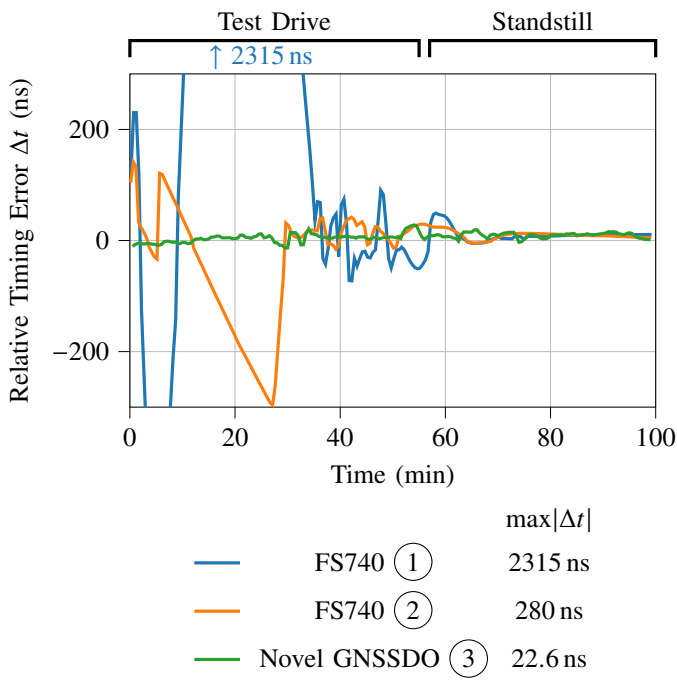
\begin{figure}[h!]
\centering
\scalebox{1}{\input{figures/MeasurementResults}}
\caption{
Measured relative timing error. When in motion, state-of-the-art devices exhibit massive instabilities.
Our novel solution, designed specifically for this use case, provides an improvement by more than an order of magnitude.
}
\label{fig:MeasurementResults}
\end{figure}

A test drive was then carried out with this setup, which covered a variety of environments:
Between high-rise buildings with severely restricted view of the sky, over rural roads, and highways.
In addition, a stress test for acceleration sensitivity was performed by moving in tight circles.
The observed differential timing errors are shown in Figure~\ref{fig:MeasurementResults}.
One of the FS740s deviated from the reference device by up to \SI{280}{\nano\second}, while the other yielded a maximum error of \SI{2315}{\nano\second}.
This significant variation in performance indicates severe instability in this environment.
All FS740s issued occasional warnings that the range of their \gls{EFC} was exceeded.
This shows that these devices have serious problems in mobile use.
In contrast, the deviation between the two new \glspl{GNSSDO} never exceeded \SI{22.6}{\nano\second}.
This constitutes an improvement of more than one order of magnitude compared to the state of the art.

\section{Conclusion}

This paper presents a synchronization solution optimized for road-mobile microwave measurements.
It was validated in a test drive with varying dynamic accelerations and \gls{GNSS} reception conditions.
This showed that classic laboratory technology has significant performance problems in this environment.
In contrast, the solution presented here reduced the relative timing error between nodes by more than an order of magnitude over the state of the art.
Further optimizations, for example, of the control algorithm, may enable even better results.

\section*{Acknowledgment}
The work was partially funded by the Federal Ministry for Economic Affairs and Climate Action in the project DOCT-Digital OTA Connectivity Twin (Förderkennzeichen: 1922001H). The hardware used in this work was partially funded by the Free State of Thuringia under grants 2015 FGI 0020, 2018 FGI 0040 and 2018 FGI 0041, and co-financed by the European Union in the frame of the European Regional Development Fund (EFRE). Furthermore, the authors acknowledge the additional financial support of
the German Federal Ministry of Education and Research in
the “Souver\"an. Digital. Vernetzt.” joint project 6G Research
Innovation Cluster (6G-RIC) (grant numbers: 16KISK020K,
16KISK030). 


\bibliographystyle{IEEEtran}
\bibliography{paper}

\end{document}

%% file: figures/BlockDiagram.tex
\begin{tikzpicture}

    \tikzset{
       boxa/.style={rectangle, draw =black,minimum width = 1.8cm, minimum height = 1.1cm,align = center}
    }

    \node [boxa](MCU){MCU};
    \node [boxa, below left = 0.5cm and 0.8cm of MCU.south east,text width=1.6cm] (TT1){Time- \\ Tagging};
    \node [boxa, below = .5cm of TT1,text width=1.6cm] (TT2){Time- \\ Tagging};
    \node [boxa, above left= .5cm and .5cm of MCU.north] (OCXO){OCXO};
    \node [boxa, right= .5cm of OCXO, text width=1.6cm] (Dist){Clock \\ Distribution};
    \node [boxa, left= .5cm of TT1,text width=1.6cm] (GNSS){GNSS \\ Module};

    \draw[->,line width = 1pt](OCXO) -- (Dist);
    \draw[->,line width = 1pt](GNSS) -- (TT1);
    \draw[->,line width = 1pt](MCU) -| (OCXO);

    \draw[->,line width = 1pt]([yshift=0.2cm]TT1.east) -| ([xshift=0.3cm]MCU.south);
    \draw[->,line width = 1pt]([yshift=0.2cm]TT2.east) -| ([xshift=0.5cm]MCU.south);

    \draw[->,line width = 1pt]([xshift=0.3cm]Dist.south) |- ([yshift=-0.2cm]TT2.east);
    \draw[->,line width = 1pt]([xshift=0.3cm]Dist.south) |- ([yshift=-0.2cm]TT1.east);
    \draw[->,line width = 1pt]([xshift=0.3cm]Dist.south) |- ([yshift=-0.2cm]MCU.east);

    \draw[->,line width = 1pt](Dist.east) -- ++( 1.4cm,0) node [midway, above, align=center, text width=1.3cm] {5x Clock \\ Export};
    \draw[->,line width = 1pt](MCU.east) -- ++( 2.4cm,0) node [pos=0.75, above, align=center, text width=1.3cm] {5x Pulse \\ Export};

    \draw[<-,line width = 1pt](TT2.west) -- ++( -1.4cm,0) node [pos=0.5, below, align=center, text width=1.3cm] {Pulse \\ Import};
    \draw[<-,line width = 1pt](GNSS.west) -- ++( -1.4cm,0) node [pos=0.5, below, align=center, text width=1.3cm] {Antenna};

\end{tikzpicture}

%% file: figures/MeasurementSetup.tex
\begin{tikzpicture}

    \definecolor{color0}{rgb}{0.12156862745098,0.466666666666667,0.705882352941177}
    \definecolor{color1}{rgb}{1,0.498039215686275,0.0549019607843137}

    \tikzset{
       box/.style={rectangle, inner sep=0.1cm, thin, draw=black,text width = 1.7cm,align=center, minimum height= 0.8cm}
    }

    \node[inner sep=0pt,opacity=0.2] (russell) at (1,-2.6)
    {\includegraphics[width=9cm,angle=-90]{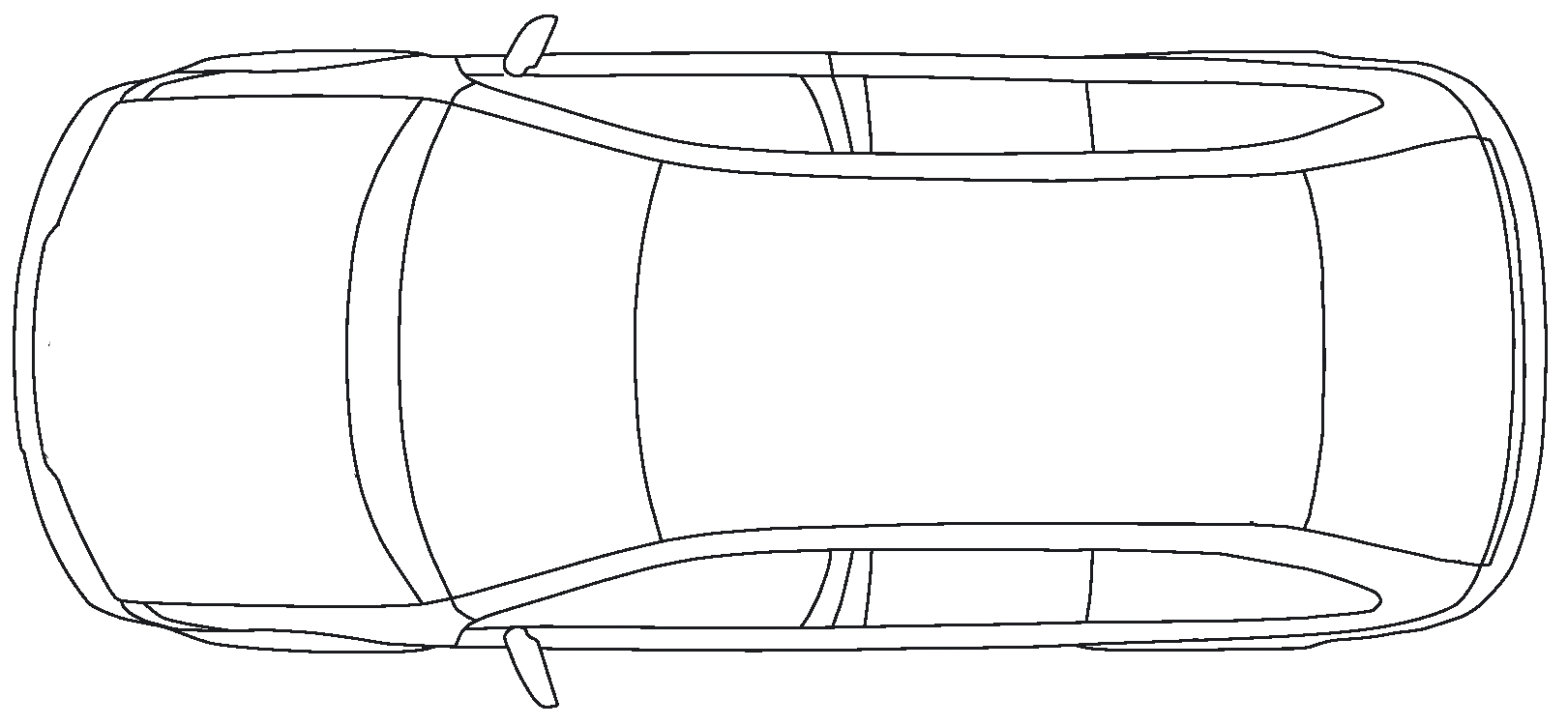}};

    \draw[rounded corners, color=color0,fill=color0!40, fill opacity=0.7] (-2.4, 1.5) rectangle (4, -3.3);
    \node[anchor=north east,font=\bfseries] at (4, 1.5) {State-of-the-Art Devices - FS740};

    \node[box] (fs_ref) at (0,0) {GNSSDO}; \draw ([yshift=0.2cm]fs_ref.west) -- ++(-0.3cm,0) node[pos=1, antenna, scale=0.3] {};
    \node[box, below=0.5cm of fs_ref, anchor=north] (fs_1){GNSSDO}; \draw ([yshift=0.2cm]fs_1.west) -- ++(-0.3cm,0) node[pos=1, antenna, scale=0.3] {};
    \node[box, below=0.5cm of fs_1, anchor=north] (fs_2){GNSSDO}; \draw ([yshift=0.2cm]fs_2.west) -- ++(-0.3cm,0) node[pos=1, antenna, scale=0.3] {};

    \draw[<->, very thick] (fs_ref.east) -- ++(0.8cm,0) |- (fs_1.east)
        node [pos=0.25, right, align=center] (mid1) {\circled{1}};
    \draw[<->, very thick] ([xshift=1.2cm]fs_ref.east) -- ++(0.8cm,0) |- ([xshift=1.2cm]fs_2.east)
        node [pos=0.25, right, align=center] (mid1) {\circled{2}};

    \draw[rounded corners, color=color1,fill=color1!40, fill opacity=0.7] (-2.4, -3.5) rectangle (4, -6.7);
    \node[anchor=north east,font=\bfseries] at (4, -3.5) {Novel Solution};

    \node[box, below=1.2cm of fs_2] (new_ref) {GNSSDO}; \draw ([yshift=0.2cm]new_ref.west) -- ++(-0.3cm,0) node[pos=1, antenna, scale=0.3] {};
    \node[box, below=0.5cm of new_ref] (new_1){GNSSDO}; \draw ([yshift=0.2cm]new_1.west) -- ++(-0.3cm,0) node[pos=1, antenna, scale=0.3] {};

    \draw[->] ([yshift=-0.2cm]fs_ref.west) -- ++(-0.8cm,0) |- ([yshift=-0.2cm]new_1.west) node [above, rotate=90, pos=0.12] {Pulse Signal};
    \draw[<-] ([yshift=-0.2cm]new_ref.west) -- ++(-0.8cm,0);
    \draw[<-] ([yshift=-0.2cm]fs_1.west) -- ++(-0.8cm,0);
    \draw[<-] ([yshift=-0.2cm]fs_2.west) -- ++(-0.8cm,0);

    \draw[<->, very thick] (new_ref.east) -- ++(0.8cm,0) |- (new_1.east)
        node [pos=0.25, right, align=center] (mid1) {\circled{3}};

\end{tikzpicture}

%% file: figures/MeasurementResults.tex
\begin{tikzpicture}
    \definecolor{color0}{rgb}{0.12156862745098,0.466666666666667,0.705882352941177}
    \definecolor{color1}{rgb}{1,0.498039215686275,0.0549019607843137}
    \definecolor{color2}{rgb}{0.172549019607843,0.627450980392157,0.172549019607843}

    \begin{axis}[
            tick align=outside,
            tick pos=left,
            xlabel={Time (min)},
            xmajorgrids,
            xtick style={color=black},
            ylabel={Relative Timing Error $\Delta t$ (ns)},
            ymajorgrids,
            xminorgrids,
            ymin=-300, ymax=300,
            xmin=0, xmax=100,
            grid=both, clip mode=individual,
            ytick style={color=black},
        ]

\draw[very thick]  (axis description cs:0,1.05)  -- (axis description cs:0,1.1) -- (axis description cs:0.55,1.1) node[midway,above] {Test Drive} -- (axis description cs:0.55,1.05)   ;
\draw[very thick]  (axis description cs:0.57,1.05)  -- (axis description cs:0.57,1.1) -- (axis description cs:1,1.1) node[midway,above] {Standstill} -- (axis description cs:1,1.05)   ;

\node[color0] at (axis description cs:0.25,1.05) {$\uparrow$ \SI{2315}{\nano\second}};

\addplot [very thick, color0] table[x=time, y=offset, row sep=crcr] {%
time offset
-0.293333 -99.608000 \\
0.206667 121.799000 \\
0.706667 230.441000 \\
1.206667 230.320000 \\
1.706667 119.667000 \\
2.206667 -124.382000 \\
2.706667 -246.589000 \\
3.206667 -368.546000 \\
3.706667 -490.419000 \\
4.206667 -612.835000 \\
4.706667 -734.651000 \\
5.206667 -778.993000 \\
5.706667 -742.531000 \\
6.206667 -631.426000 \\
6.706667 -520.763000 \\
7.206667 -410.017000 \\
7.706667 -299.327000 \\
8.206667 -212.189000 \\
8.706667 -142.952000 \\
9.206667 77.722000 \\
9.706667 188.583000 \\
10.206667 299.199000 \\
10.706667 409.813000 \\
11.206667 520.370000 \\
11.706667 631.172000 \\
12.206667 741.764000 \\
12.706667 852.446000 \\
13.206667 963.116000 \\
13.706667 1073.637000 \\
14.206667 1184.676000 \\
14.706667 1295.108000 \\
15.206667 1406.154000 \\
15.706667 1516.848000 \\
16.206667 1627.784000 \\
16.706667 1738.463000 \\
17.206667 1849.605000 \\
17.706667 1960.795000 \\
18.206667 2071.652000 \\
18.706667 2182.728000 \\
19.206667 2293.757000 \\
19.706667 2404.517000 \\
20.206667 2515.452000 \\
20.706667 2546.318000 \\
21.206667 2460.778000 \\
21.706667 2338.993000 \\
22.206667 2217.059000 \\
22.706667 2095.135000 \\
23.206667 1973.260000 \\
23.706667 1851.318000 \\
24.206667 1729.700000 \\
24.706667 1607.626000 \\
25.206667 1485.785000 \\
25.706667 1363.761000 \\
26.206667 1241.903000 \\
26.706667 1119.997000 \\
27.206667 998.074000 \\
27.706667 924.804000 \\
28.206667 877.020000 \\
28.706667 828.259000 \\
29.206667 775.516000 \\
29.706667 718.087000 \\
30.206667 654.963000 \\
30.706667 584.521000 \\
31.206667 520.214000 \\
31.706667 452.720000 \\
32.206667 387.841000 \\
32.706667 327.574000 \\
33.206667 271.068000 \\
33.706667 217.650000 \\
34.206667 164.542000 \\
34.706667 103.282000 \\
35.206667 36.280000 \\
35.706667 69.969000 \\
36.206667 68.453000 \\
36.706667 -32.534000 \\
37.206667 -42.588000 \\
37.706667 -14.311000 \\
38.206667 25.565000 \\
38.706667 40.815000 \\
39.206667 47.712000 \\
39.706667 -28.825000 \\
40.206667 64.033000 \\
40.706667 74.214000 \\
41.206667 30.715000 \\
41.706667 -72.036000 \\
42.206667 -72.683000 \\
42.706667 -32.205000 \\
43.206667 -45.888000 \\
43.706667 -32.746000 \\
44.206667 -20.348000 \\
44.706667 -19.323000 \\
45.206667 -29.912000 \\
45.706667 -31.251000 \\
46.206667 -21.411000 \\
46.706667 -16.613000 \\
47.206667 34.654000 \\
47.706667 90.328000 \\
48.206667 82.502000 \\
48.706667 29.333000 \\
49.206667 -31.785000 \\
49.706667 -49.312000 \\
50.206667 -32.341000 \\
50.706667 -16.926000 \\
51.206667 -16.592000 \\
51.706667 -11.868000 \\
52.206667 -17.059000 \\
52.706667 -26.436000 \\
53.206667 -36.000000 \\
53.706667 -43.258000 \\
54.206667 -48.872000 \\
54.706667 -50.841000 \\
55.206667 -49.389000 \\
55.706667 -42.831000 \\
56.206667 -34.968000 \\
56.706667 -20.038000 \\
57.206667 32.109000 \\
57.706667 46.502000 \\
58.206667 49.389000 \\
58.706667 46.953000 \\
59.206667 45.823000 \\
59.706667 44.858000 \\
60.206667 41.332000 \\
60.706667 35.305000 \\
61.206667 27.357000 \\
61.706667 18.561000 \\
62.206667 9.985000 \\
62.706667 5.728000 \\
63.206667 4.373000 \\
63.706667 1.490000 \\
64.206667 -2.707000 \\
64.706667 -4.288000 \\
65.206667 -5.069000 \\
65.706667 -5.128000 \\
66.206667 -4.934000 \\
66.706667 -4.634000 \\
67.206667 -4.161000 \\
67.706667 -3.170000 \\
68.206667 -1.851000 \\
68.706667 1.590000 \\
69.206667 3.475000 \\
69.706667 5.552000 \\
70.206667 5.870000 \\
70.706667 3.551000 \\
71.206667 3.566000 \\
71.706667 3.511000 \\
72.206667 3.246000 \\
72.706667 3.271000 \\
73.206667 7.394000 \\
73.706667 7.638000 \\
74.206667 7.734000 \\
74.706667 3.818000 \\
75.206667 3.721000 \\
75.706667 7.490000 \\
76.206667 7.585000 \\
76.706667 7.584000 \\
77.206667 7.623000 \\
77.706667 7.755000 \\
78.206667 7.751000 \\
78.706667 7.846000 \\
79.206667 7.817000 \\
79.706667 7.881000 \\
80.206667 8.058000 \\
80.706667 8.199000 \\
81.206667 8.347000 \\
81.706667 8.577000 \\
82.206667 8.785000 \\
82.706667 9.106000 \\
83.206667 9.261000 \\
83.706667 9.402000 \\
84.206667 9.548000 \\
84.706667 9.626000 \\
85.206667 9.642000 \\
85.706667 9.628000 \\
86.206667 9.603000 \\
86.706667 9.708000 \\
87.206667 9.695000 \\
87.706667 9.794000 \\
88.206667 9.930000 \\
88.706667 10.009000 \\
89.206667 10.263000 \\
89.706667 10.422000 \\
90.206667 10.497000 \\
90.706667 10.658000 \\
91.206667 10.648000 \\
91.706667 10.635000 \\
92.206667 10.663000 \\
92.706667 10.698000 \\
93.206667 10.721000 \\
93.706667 10.742000 \\
94.206667 10.693000 \\
94.706667 10.696000 \\
95.206667 10.604000 \\
95.706667 10.599000 \\
96.206667 10.600000 \\
96.706667 10.541000 \\
97.206667 10.619000 \\
97.706667 10.679000 \\
98.206667 10.756000 \\
98.706667 10.853000 \\
99.206667 10.901000 \\
}; \label{plot:sota1}

\addplot [very thick, color1] table[x=time, y=offset, row sep=crcr] {%
time offset
-0.341667 -64.318000 \\
0.158333 104.241000 \\
0.658333 141.150000 \\
1.158333 133.415000 \\
1.658333 33.132000 \\
2.158333 23.408000 \\
2.658333 14.690000 \\
3.158333 5.453000 \\
3.658333 -11.599000 \\
4.158333 -20.972000 \\
4.658333 -29.453000 \\
5.158333 -33.196000 \\
5.658333 121.133000 \\
6.158333 119.085000 \\
6.658333 108.372000 \\
7.158333 98.155000 \\
7.658333 88.171000 \\
8.158333 77.741000 \\
8.658333 68.064000 \\
9.158333 57.970000 \\
9.658333 47.359000 \\
10.158333 37.048000 \\
10.658333 26.895000 \\
11.158333 16.609000 \\
11.658333 5.960000 \\
12.158333 -14.107000 \\
12.658333 -24.140000 \\
13.158333 -34.470000 \\
13.658333 -44.218000 \\
14.158333 -54.449000 \\
14.658333 -64.882000 \\
15.158333 -74.868000 \\
15.658333 -85.112000 \\
16.158333 -95.300000 \\
16.658333 -105.099000 \\
17.158333 -114.921000 \\
17.658333 -125.049000 \\
18.158333 -134.945000 \\
18.658333 -144.854000 \\
19.158333 -154.865000 \\
19.658333 -164.755000 \\
20.158333 -175.043000 \\
20.658333 -185.474000 \\
21.158333 -194.155000 \\
21.658333 -203.252000 \\
22.158333 -211.791000 \\
22.658333 -220.928000 \\
23.158333 -229.757000 \\
23.658333 -238.883000 \\
24.158333 -247.906000 \\
24.658333 -256.778000 \\
25.158333 -265.612000 \\
25.658333 -274.368000 \\
26.158333 -283.230000 \\
26.658333 -291.951000 \\
27.158333 -296.137000 \\
27.658333 -259.245000 \\
28.158333 -194.196000 \\
28.658333 -129.594000 \\
29.158333 -69.016000 \\
29.658333 31.510000 \\
30.158333 27.088000 \\
30.658333 15.914000 \\
31.158333 7.975000 \\
31.658333 5.860000 \\
32.158333 3.619000 \\
32.658333 5.982000 \\
33.158333 10.819000 \\
33.658333 15.584000 \\
34.158333 18.345000 \\
34.658333 19.165000 \\
35.158333 17.927000 \\
35.658333 17.330000 \\
36.158333 39.702000 \\
36.658333 40.528000 \\
37.158333 17.294000 \\
37.658333 -7.840000 \\
38.158333 -5.768000 \\
38.658333 -8.369000 \\
39.158333 -15.620000 \\
39.658333 -10.720000 \\
40.158333 13.418000 \\
40.658333 25.259000 \\
41.158333 35.799000 \\
41.658333 39.122000 \\
42.158333 42.036000 \\
42.658333 37.582000 \\
43.158333 27.104000 \\
43.658333 30.624000 \\
44.158333 33.830000 \\
44.658333 24.645000 \\
45.158333 -7.561000 \\
45.658333 -5.083000 \\
46.158333 11.762000 \\
46.658333 24.944000 \\
47.158333 24.785000 \\
47.658333 16.529000 \\
48.158333 13.138000 \\
48.658333 5.019000 \\
49.158333 2.414000 \\
49.658333 -5.632000 \\
50.158333 -13.716000 \\
50.658333 -11.118000 \\
51.158333 6.956000 \\
51.658333 9.537000 \\
52.158333 12.400000 \\
52.658333 15.412000 \\
53.158333 19.147000 \\
53.658333 23.901000 \\
54.158333 26.394000 \\
54.658333 27.491000 \\
55.158333 29.112000 \\
55.658333 29.485000 \\
56.158333 29.187000 \\
56.658333 27.558000 \\
57.158333 25.782000 \\
57.658333 24.350000 \\
58.158333 24.021000 \\
58.658333 24.098000 \\
59.158333 24.065000 \\
59.658333 23.872000 \\
60.158333 23.252000 \\
60.658333 22.048000 \\
61.158333 20.085000 \\
61.658333 17.454000 \\
62.158333 14.571000 \\
62.658333 11.425000 \\
63.158333 8.356000 \\
63.658333 4.812000 \\
64.158333 -1.517000 \\
64.658333 -3.388000 \\
65.158333 -4.412000 \\
65.658333 -4.631000 \\
66.158333 -4.597000 \\
66.658333 -4.142000 \\
67.158333 -3.539000 \\
67.658333 -2.907000 \\
68.158333 -2.041000 \\
68.658333 -1.269000 \\
69.158333 1.944000 \\
69.658333 4.048000 \\
70.158333 6.374000 \\
70.658333 8.338000 \\
71.158333 9.878000 \\
71.658333 10.893000 \\
72.158333 11.368000 \\
72.658333 11.844000 \\
73.158333 12.298000 \\
73.658333 12.547000 \\
74.158333 12.707000 \\
74.658333 12.826000 \\
75.158333 12.889000 \\
75.658333 12.884000 \\
76.158333 12.781000 \\
76.658333 12.666000 \\
77.158333 12.637000 \\
77.658333 12.432000 \\
78.158333 12.290000 \\
78.658333 12.204000 \\
79.158333 12.045000 \\
79.658333 11.821000 \\
80.158333 11.623000 \\
80.658333 11.557000 \\
81.158333 11.345000 \\
81.658333 11.095000 \\
82.158333 10.976000 \\
82.658333 10.662000 \\
83.158333 10.504000 \\
83.658333 10.337000 \\
84.158333 10.227000 \\
84.658333 10.133000 \\
85.158333 9.929000 \\
85.658333 9.666000 \\
86.158333 9.498000 \\
86.658333 9.268000 \\
87.158333 9.133000 \\
87.658333 8.827000 \\
88.158333 8.706000 \\
88.658333 8.641000 \\
89.158333 8.638000 \\
89.658333 8.617000 \\
90.158333 8.531000 \\
90.658333 8.352000 \\
91.158333 8.116000 \\
91.658333 8.001000 \\
92.158333 7.712000 \\
92.658333 7.513000 \\
93.158333 7.370000 \\
93.658333 7.239000 \\
94.158333 6.951000 \\
94.658333 6.838000 \\
95.158333 6.483000 \\
95.658333 6.310000 \\
96.158333 6.111000 \\
96.658333 5.964000 \\
97.158333 5.722000 \\
97.658333 5.567000 \\
98.158333 5.523000 \\
98.658333 5.363000 \\
99.158333 5.146000 \\
}; \label{plot:sota2}

\addplot [very thick, color2] table[x=time, y=offset, row sep=crcr] {%
time offset
0.000000 -4.180447 \\
0.500000 -10.556617 \\
1.000000 -7.216043 \\
1.500000 -5.134882 \\
2.000000 -4.911250 \\
2.500000 -6.183357 \\
3.000000 -5.947599 \\
3.501667 -5.438222 \\
4.003333 -6.080563 \\
4.503333 -6.811187 \\
5.003333 -7.473335 \\
5.503333 -8.567899 \\
6.003333 -7.135485 \\
6.503333 -2.158621 \\
7.003333 -1.117660 \\
7.503333 -1.572684 \\
8.003333 -3.074659 \\
8.503333 -2.828911 \\
9.003333 -3.926634 \\
9.503333 -5.861354 \\
10.003333 -3.199651 \\
10.503333 -3.559013 \\
11.003333 -5.599651 \\
11.503333 -5.071965 \\
12.003333 -1.017031 \\
12.503333 2.699222 \\
13.003333 2.655813 \\
13.503333 4.141952 \\
14.003333 1.706123 \\
14.503333 1.549976 \\
15.003333 3.050499 \\
15.503333 6.152500 \\
16.003333 6.299788 \\
16.503333 8.134492 \\
17.003333 8.021116 \\
17.503333 4.974987 \\
18.003333 4.719570 \\
18.503333 4.992754 \\
19.003333 5.116611 \\
19.503333 3.205524 \\
20.003333 3.692063 \\
20.503333 4.210815 \\
21.005000 9.739882 \\
21.505000 9.771513 \\
22.005000 5.788887 \\
22.505000 5.466229 \\
23.005000 3.897235 \\
23.505000 6.477101 \\
24.005000 7.421593 \\
24.505000 11.561029 \\
25.005000 12.053946 \\
25.505000 4.993611 \\
26.005000 5.887303 \\
26.506667 8.802284 \\
27.006667 5.109608 \\
27.506667 -3.782657 \\
28.006667 -4.199013 \\
28.506667 -8.750357 \\
29.006667 -8.801464 \\
29.506667 -13.613205 \\
30.006667 -13.303655 \\
30.506667 14.251736 \\
31.006667 14.314385 \\
31.506667 14.104463 \\
32.006667 8.844624 \\
32.506667 -7.635321 \\
33.006667 -7.604044 \\
33.508333 12.917557 \\
34.008333 21.024507 \\
34.508333 12.081247 \\
35.008333 11.456980 \\
35.508333 10.488349 \\
36.008333 6.951439 \\
36.508333 6.673588 \\
37.008333 5.546909 \\
37.508333 5.235082 \\
38.008333 3.980147 \\
38.508333 4.193827 \\
39.008333 6.197773 \\
39.508333 6.426430 \\
40.008333 6.149607 \\
40.508333 6.742409 \\
41.008333 7.744200 \\
41.508333 7.574434 \\
42.008333 6.114595 \\
42.508333 5.970499 \\
43.008333 4.287661 \\
43.508333 6.597290 \\
44.008333 6.398420 \\
44.508333 7.024720 \\
45.008333 3.849412 \\
45.508333 1.675549 \\
46.008333 3.857993 \\
46.508333 5.688889 \\
47.008333 4.697991 \\
47.508333 3.196808 \\
48.008333 4.721677 \\
48.508333 5.613400 \\
49.008333 5.274723 \\
49.508333 6.515524 \\
50.008333 6.223760 \\
50.508333 5.055170 \\
51.008333 7.047413 \\
51.508333 3.620392 \\
52.008333 15.168003 \\
52.508333 22.024469 \\
53.008333 24.285845 \\
53.508333 26.613091 \\
54.008333 28.080435 \\
54.508333 26.697640 \\
55.008333 16.208345 \\
55.508333 6.740156 \\
56.008333 1.825248 \\
56.508333 5.824561 \\
57.008333 9.773696 \\
57.508333 8.453211 \\
58.008333 9.323089 \\
58.508333 10.642713 \\
59.008333 9.575693 \\
59.508333 6.692944 \\
60.008333 6.654388 \\
60.508333 7.273730 \\
61.008333 6.719818 \\
61.508333 4.329755 \\
62.008333 4.157841 \\
62.508333 -1.306833 \\
63.008333 10.286758 \\
63.508333 15.449188 \\
64.008333 15.391205 \\
64.508333 15.076609 \\
65.008333 15.032022 \\
65.508333 19.240688 \\
66.008333 19.453777 \\
66.508333 13.602489 \\
67.008333 9.662531 \\
67.508333 6.736586 \\
68.008333 2.540491 \\
68.508333 9.698763 \\
69.008333 11.410402 \\
69.508333 10.875711 \\
70.008333 8.563339 \\
70.508333 10.959838 \\
71.008333 13.082767 \\
71.508333 13.790176 \\
72.008333 14.667336 \\
72.508333 14.875723 \\
73.008333 14.741351 \\
73.508333 7.291468 \\
74.008333 -3.228070 \\
74.508333 -2.254778 \\
75.008333 2.766170 \\
75.508333 7.533892 \\
76.008333 10.271795 \\
76.508333 10.387333 \\
77.008333 10.069480 \\
77.508333 7.920941 \\
78.008333 4.369108 \\
78.508333 2.950954 \\
79.008333 3.074885 \\
79.508333 4.300169 \\
80.008333 9.271384 \\
80.508333 10.173745 \\
81.008333 9.762695 \\
81.508333 10.014204 \\
82.008333 9.437841 \\
82.508333 10.054915 \\
83.008333 11.869917 \\
83.508333 11.529814 \\
84.008333 12.543213 \\
84.508333 12.565213 \\
85.008333 8.548247 \\
85.508333 9.255045 \\
86.008333 10.495706 \\
86.508333 10.881340 \\
87.008333 12.092267 \\
87.508333 10.515478 \\
88.008333 9.899467 \\
88.508333 10.603247 \\
89.008333 11.268528 \\
89.508333 12.017197 \\
90.008333 12.646467 \\
90.508333 13.743519 \\
91.008333 15.863738 \\
91.508333 15.589783 \\
92.008333 15.893126 \\
92.508333 15.077455 \\
93.008333 12.159452 \\
93.508333 11.193336 \\
94.008333 10.323029 \\
94.508333 12.706491 \\
95.008333 14.010242 \\
95.508333 12.660709 \\
96.008333 8.452834 \\
96.508333 4.989811 \\
97.008333 4.076384 \\
97.508333 3.514620 \\
98.008333 1.851307 \\
98.508333 1.913174 \\
99.008333 1.259161 \\
}; \label{plot:novel}
\coordinate (legend) at (axis description cs:0.5,-0.25);

    \end{axis}

            \matrix [
            matrix of nodes,
            anchor=north, row sep=-0.1cm
        ] at (legend) {
                             &     & $\max \lvert \Delta t \rvert$ \\
            \ref{plot:sota1} & FS740 \circled{1}   & \SI{2315}{\nano\second}   \\
            \ref{plot:sota2} & FS740 \circled{2}   & \SI{280}{\nano\second}   \\
            \ref{plot:novel} & Novel GNSSDO \circled{3}  & \SI{22.6}{\nano\second}   \\
        };
\end{tikzpicture}